\documentclass{aa}
\usepackage{graphicx,subfigure,amssymb}
\def\es{erg~s$^{-1}$}

\def\xmm{{\it XMM-Newton\/}}
\def\etal{et al.\ }
\def\betamod{$\beta$-model}

\def\Lx {L_{\rm X}}
\def\Tx {T}

\def \h50 {$h_{50}$}
\def \h75 {$h_{75}$}

\def\es{erg~s$^{-1}$}

\begin{document}

\title{An \xmm\ observation of A3921: an off-axis merger}

\author{E. Belsole\inst{1,2} \and J-L. Sauvageot\inst{1} \and G.W. Pratt\inst{1,3}  \and H. Bourdin\inst{4,5}}
\offprints{E. Belsole,\\email:e.belsole@bristol.ac.uk}

\institute{$^1$Service d'Astrophysique, CEA Saclay, L'Orme des Merisiers 
B\^at 709., F-91191 Gif-sur-Yvette Cedex, France\\
\email{jsauvageot@cea.fr}\\
$^2$H.H. Wills Physics Laboratory, University of Bristol, Tyndall Avenue, Bristol BS8 1TL, U.K.\\
$^3$Max-Planck-Institut f\"ur extraterrestrische Physik, Postfach 1312, 85741 Garching, Germany\\
\email{gwp@mpe.mpg.de}\\
$^4$Observatoire de la C\^ote d'Azur, BP 4229, F-06304 Nice Cedex 4, France \\
$^5$Dipartimento di Fisica, Universit\'a degli Studi di Roma "Tor Vergata", Via della Ricerca Scientifica, 1, 00133 Roma, Italy\\
\email{Herve.Bourdin@roma2.infn.it}
}

\date{Received ; accepted}

\abstract{We present the results of a detailed analysis of the \xmm\ observation of the galaxy cluster Abell 3921. The X-ray morphology of the cluster is elliptical, with the centroid offset from the brightest cluster galaxy by 17\arcsec, and with a pronounced extension toward the NW. Subtraction of a 2D \betamod\ fit to the main cluster emission reveals a large scale, irregular residual structure in the direction of the extension, containing both diffuse emission from the intra cluster medium, and extended emission from the second and third-brightest cluster galaxies (BG2 and BG3). The greatest concentration of galaxies in the subcluster lies at the extreme northern edge of the residual. The cluster exhibits a remarkable temperature structure, in particular a bar of significantly hotter gas, oriented SE-NW and stretching from the centre of the cluster towards BG2 and BG3. Our detailed study of the morphological and thermal structure points to an off-axis merger between a main cluster and a less massive galaxy cluster infalling from the SE. From comparison of the temperature map with numerical simulations, and with independent calculations based on simple physical assumptions, we conclude that the merging event is $\sim 0.5$ Gyr old. The cluster is thus perhaps the best X-ray observed candidate so
far of an intermediate mass ratio, moderate impact parameter merger.
}
\maketitle

\section{Introduction}
In the current scenario for the formation of structure in the Universe, clusters of galaxies form hierarchically by a succession of mergers with objects already formed. During a merger event more than 10$^{63}$ ergs are dissipated in the intra-cluster medium (ICM) through shock waves, which form initially between the colliding objects and then propagate in the impact direction. These merger events have a strong effect on the cluster, triggering star formation (Evrard \cite{evrard91}) and gas stripping (Fujita \etal \cite{fujita99}) in the galaxies, and leading to strong variations of the physical properties (such as the temperature, the density and the entropy) of the ICM.

From the theoretical point of view, numerical N-body simulations have shown that mergers produce substructure observable in the density and temperature distribution of the ICM (e.g., Roettiger, Burns \& Loken ~\cite{roettiger96}; Ricker 1998). Mergers have also been found to modify the member galaxy distribution (Schindler \& B\"ohringer \cite{sc93}). Thus, substructures (in the ICM or in the galaxy distribution) in clusters are a fossil record of the merger history.

X-ray observations of galaxy clusters (Forman \& Jones~\cite{FJ82}; Mohr, Fabricant \& Geller~\cite{MFG93}) suggest that such objects at low redshift are dynamically young (Buote \& Tsai \cite{BuTs}), displaying a large degree of density and temperature structure (e.g., Henry \& Briel \cite{hb96}; Markevitch \& Vikhlinin \cite{MV01}; Mazzotta, Edge \& Markevitch \cite{mem03}; Neumann et al. \cite{neumancoma03}). Since the temperature substructure survives for a longer time than density substructure  (typically 4 to 6 times longer), the temperature is a strong  indicator of the cluster history and its present dynamical state.

Statistical studies of  cluster morphology can thus provide an important test of cosmological models of structure formation. The comparison of numerical simulations of structure formation to statistical studies of nearby clusters has been until now limited by the knowledge of the effect of cluster formation and evolution on the observational properties. Despite the large amount of data available, the dynamical evolution of the ICM and the relation between the galaxies and gas during a merger event is still poorly understood (see Buote \cite{buote02} for a review).  A better understanding of the physics of merger events, in particular the relaxation time of substructures (Nakamura \cite{nakamura95}) is required.

Combined X-ray and optical studies (e.g. Roettiger \etal \cite{roettiger98}; Durret \etal \cite{durret}; Ferrari \etal \cite{f03}) have the advantage of combining information on the collisional component (the gas), and the nominally non-collisional -- and thus more likely to trace the dark matter -- component (the galaxies). Such studies are thus well suited to investigate in deeper detail the dynamical process of merging clusters.

In this paper we present the results obtained from the \xmm\ observation of the merging cluster A3921. This is the second target from our \xmm\ Guaranteed  Time (GT) program, which was established as a systematic study of merging clusters of galaxies, with the aim of deepening our understanding of the process of cluster formation and evolution. 

A3921 is classified as a regular cluster of richness 2 (Abell, Corwin \& Olowin \cite{aco89}) at redshift $z=0.094$. It was detected first in X-ray by the A1 survey of HEAO-1 (Kolvalski \etal~\cite{kolvalski84}) and was subsequently observed in pointed observations by the {\em Einstein}, ROSAT and GINGA satellites. The latter two observations have been analysed by Arnaud et al.~(\cite{arnauda3921}) who showed that A3921 consists of two merging components with a mass ratio of 1 to 3. Their analysis indicated that the smaller subcluster is located at a distance of 0.8 \h75 $^{-1}$ Mpc  to the West of the main cluster. The ROSAT/GINGA observation indicated that the region between the subcluster and the main cluster has a higher temperature, however, the poor spatial resolution and limited energy coverage of ROSAT prevented deeper interpretation. A3921 was also observed with ASCA. White (\cite{white2000}) measured a global temperature of 5.73$_{-0.23}^{+0.24}$ keV and a strong cooling flow of mass deposition rate $\dot{M} = 319$ M$_{\odot}$ yr$^{-1}$. The temperature profile obtained by White (\cite{white2000}) is fairly isothermal, with an indication of a higher temperature at the edge of the cluster. However, the spatial resolution of the satellite did not allow a precise analysis. All published indications thus point to a merging cluster in an intermediate phase of the merger, just before the core passage.  It was on the basis of this information that A3921 was selected for our sample of merging clusters of galaxies. 

In this paper we use the high sensitivity and good spatial resolution of \xmm\ to gain new insights into this system. Our conclusions are the result of a comparison of our X-ray observation with numerical simulations and the new multi-object spectroscopy and VRI-band imaging data obtained at the ESO telescopes and detailed in an accompanying paper (Ferrari~\etal~2004; hereafter F04), and allow us to build an entirely new picture of the dynamical state of the system. 

Throughout the paper we assume $H_0 = 75$ km s$^{-1}$ Mpc$^{-1}$, $\Omega_m = 0.3$, $\Omega_{\Lambda}$ = 0.7.  In this cosmology, at the redshift of the cluster, 1 arcmin corresponds to 98 kpc . 
 
\section{Observations and data preparation}

\subsection{Observations}
A3921  was observed for 30 ks in October 2000  by \xmm\ as a Guaranteed Time target. In this paper only data from the European Photon Imaging Camera (EPIC; Str\"uder et al.~\cite{strueder}; Turner et al. 2001) are considered.
Calibrated event list files were provided by the \xmm\ SOC the 19 of June 2001. All observations were obtained with the MEDIUM filter in Full Frame mode. Throughout this analysis  single pixel events for the pn data (PATTERN 0) were selected, while for the MOS data sets the PATTERNs 0-12 were used.

In order to estimate the background level and its variability with time because of flares, we produced light curves in 100s bins in a high energy band (10.0-12.0 keV for MOS and 12.0-14.-0 keV for pn). The mean level of the light curve was very regular, thus there are no flares and for the scientific analysis, we used the total exposure time of the observation, which was 30073.4 s, 30219.1 s, and 22398.1 s for MOS1, MOS2, and PN, respectively.

The background estimate was obtained using the blank-sky event lists described in Lumb \etal (\cite{lumbbkg}). The blank-sky background events were selected using the same selection criteria (such as PATTERN, FLAG, etc.) used for the observation events and transformed in order to have the same sky coordinates (EPOCH J2000) as for A3921. This procedure ensures that the source and background products come from the same region of the detector, reducing errors induced by the detector position dependence. By generating the light curve of the blank-sky background in the high energy (10-12 keV) band, we rejected time intervals displaying more than 15 counts per 100 s bin in order to have the same statistical flare-induced background as A3921. 

The source and background events were corrected for vignetting using the SAS task {\sc evigweight}, which implements the method outlined in  Briel \& Henry \cite{bh94}). This allows us to use the on-axis response matrices and effective areas.

To take into account the variation of the particle induced background,  which dominates at high energy ($>$ 5 keV) and induces fluorescence lines (Al, Si, Cu, Au) from the shielding of the camera and the detector itself, we computed normalisation factors by considering the photon (particle) statistics in the 10-12 keV (MOS) and 12-14 keV band (pn) between the source  and the blank-sky background observations. The normalisations used in this work are 0.99, 1.00, 1.12 for MOS1, MOS2 and pn respectively.

\section{The cluster X-ray morphology}
\begin{figure*}[]
\centering
\subfigure[] 
{
    \label{fig:fig1a}
    \includegraphics[scale=0.45,angle=0,keepaspectratio]{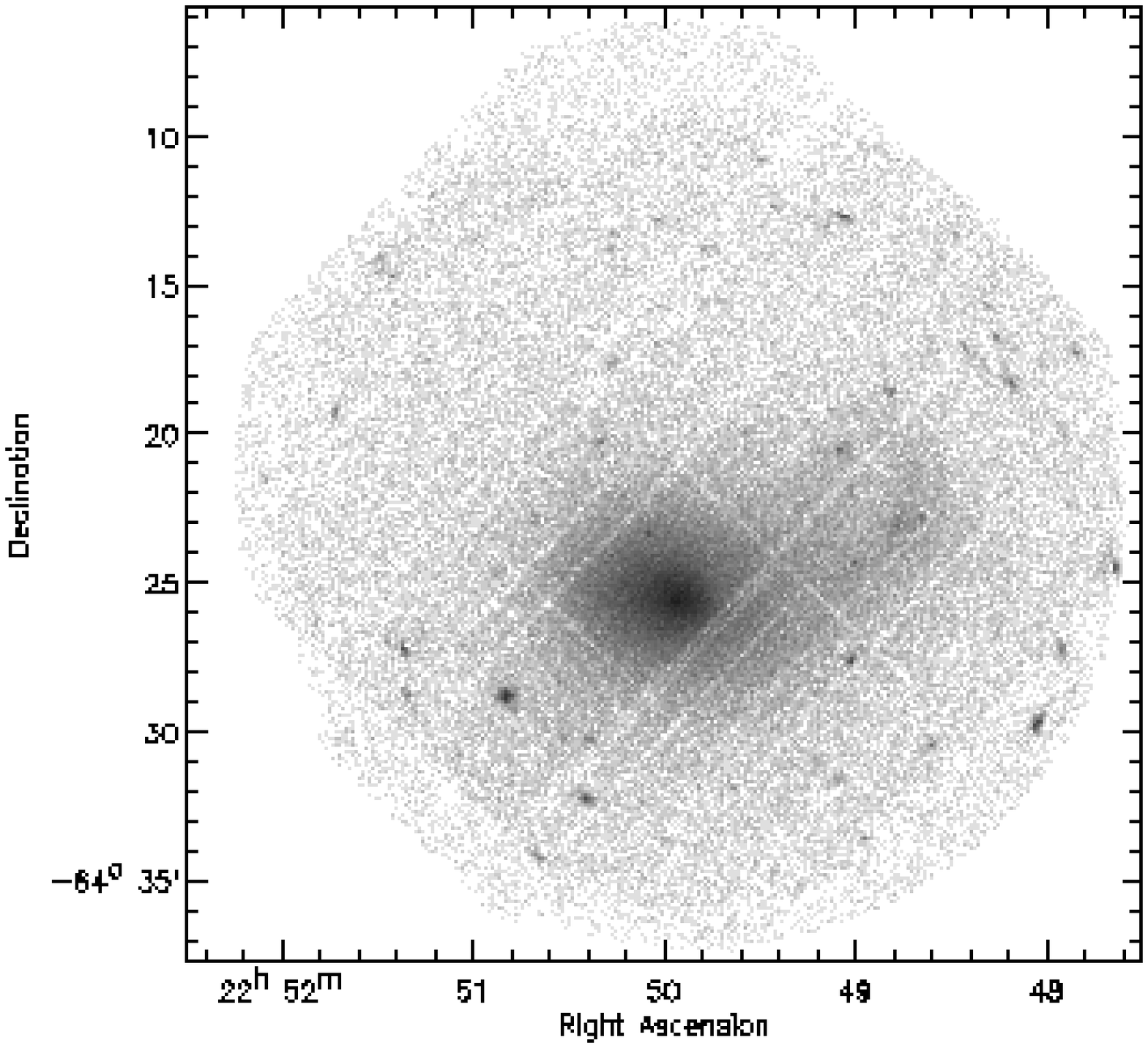}
}
\hspace{0cm}
\subfigure[] 
{
    \label{fig:fig1b}
    \includegraphics[scale=0.50,angle=0,keepaspectratio]{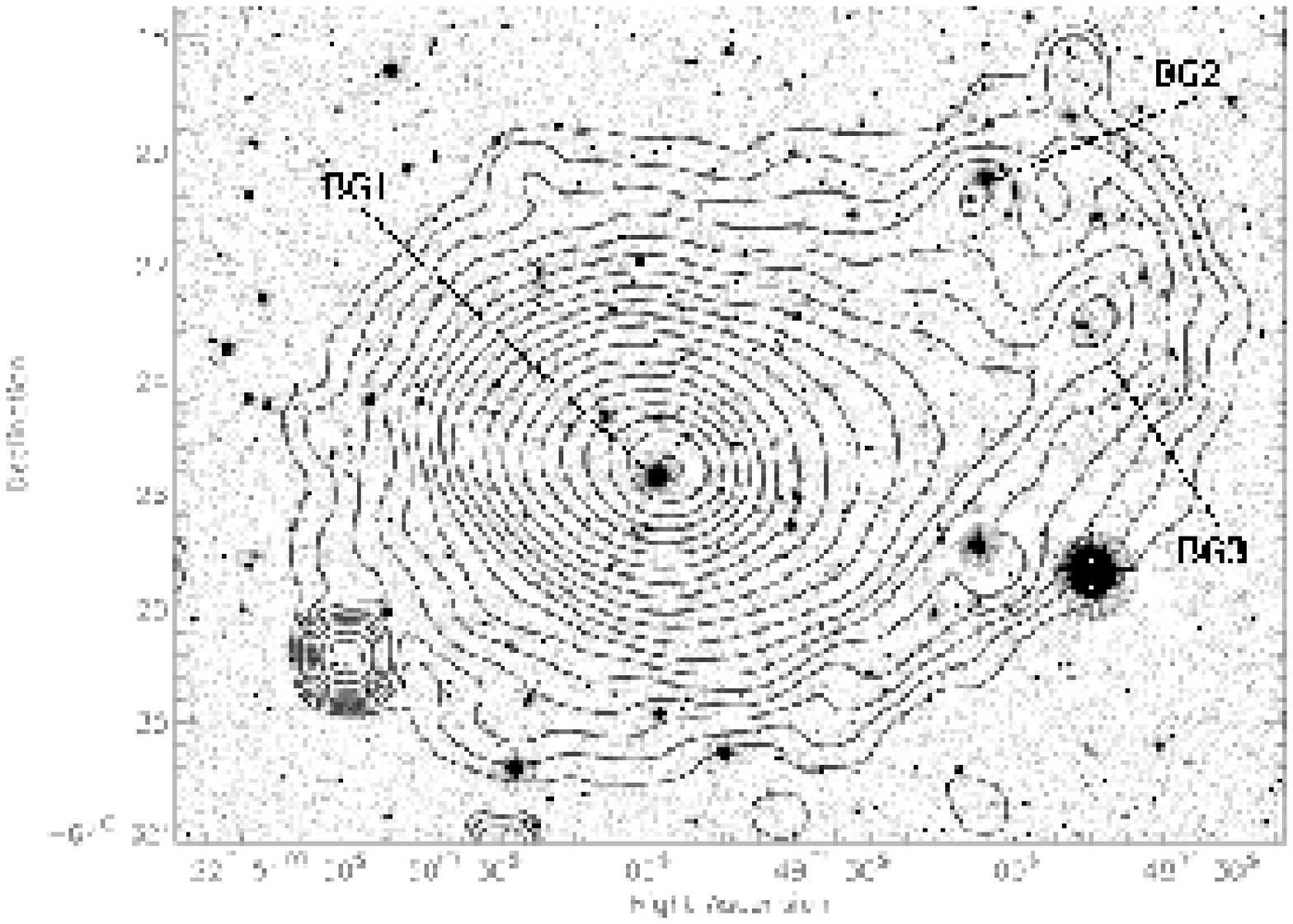}
}
\caption{(a) EPIC counts image in the energy band 0.3-10.0 keV; the image is not corrected for exposure. (b) Iso-intensity contours of the EMOS1+EMOS2 adaptively smoothed image in the energy band  0.3-10.0 keV overlaid on the deep I band optical image obtained at the MPG/ESO 2.2 m telescope (see \cite{f04}) in la Silla. The brightest galaxies are marked. Contours come from a non-background subtracted, non-exposure-corrected image, and are logarithmically spaced. The lowest contour level is at 0.6 (MOS1+MOS2) counts/pixel. Epoch is J2000.}
\label{fig:fig1} 
\end{figure*}

\subsection{X-ray image}

Figure \ref{fig:fig1a} shows the EPIC image in the 30\arcmin~ field of view (FoV) of \xmm\, obtained by combining the  MOS1, MOS2 and pn images in the energy range 0.3-10.0 keV. The image is not corrected for exposure.
The cluster displays a slightly elliptical shape, with an extension to the west-north-west (WNW). The extension displays two extended emission peaks which correspond to two of the three brightest galaxies in optical, as shown in Fig.~\ref{fig:fig1b}. Here the X-ray iso-intensity contours of the \xmm\ MOS adaptively smoothed image are superimposed on the deep I-band image obtained with the WFI instrument at the MPG/ESO 2.2 m telescope (\cite{f04}, for details). The two galaxies are at 7\arcmin~ ($690 h_{75}^{-1}$ kpc; RA = $22^h48^m49^s$, Dec = $-64^o23\arcmin10\arcsec$ (J2000)) to the north-west (BG2) and at 8\arcmin~ ($800 h_{75}^{-1}$ kpc; RA = $22^h49^m04^s$, Dec = $-64^o20\arcmin35\arcsec$ (J2000)) to the WNW (BG3) from the central brightest galaxy BG1 (RA = $22^h49^m58^s$; Dec=$-64^o25\arcmin46\arcsec$ (J2000)). The peak of the X-ray emission is shifted by 17\arcsec\ to the NW with respect to the optical  position of BG1. This is larger than any attitude error of \xmm, and note that the X-ray and optical emission of BG3 coincide well within 17\arcsec, although the X-ray emission around BG3 is extended. The peak of the BG2 X-ray emission is also shifted, this time towards the SE, by 26\arcsec. The diffuse X-ray  emission detected around galaxy BG3 corresponds to the emission of the subcluster detected by Arnaud et al. (\cite{arnauda3921}) with the ROSAT data. We also notice an interesting depression in the surface brightness, between BG2 and BG3. 

Several point and extended sources are also detected in the field of A3921. Some of them are the X-ray counterparts of galaxies in the cluster, others are unrelated objects. All point sources with flux greater than 10$^{-14}$ ergs s$^{-1}$  arcmin$^{-2}$  were masked throughout this analysis by checking the source detection file in the pipeline products. Such point sources were masked with a typical radius of 30\arcsec. Some sources marginally detected by the SAS detection algorithm were also masked.

\subsection{2D $\beta$ model fitting}\label{sec:2Dbeta}

For this analysis we have followed the method applied to the merging cluster A1750, described in Belsole et al.~(\cite{a1750}). We fit the surface brightness distribution of A3921 at low energy (where the distribution is less temperature-dependent) with a $\beta$ model, and quantify the deviation from this model. We use only the MOS camera because the large gaps in the pn camera are difficult to take into account, and could produce spurious structure.

In order for the model to be a good representation of the main cluster, we masked the region connected to the north-western extension with a rectangle of size $10.3 \times 13.0$ arcmin, together with the mask used to excise point sources. The centre of the cluster was also masked with a circle of 1\farcm36 (the results are identical whether or not the central region is excluded from the fit). The best-fitting 2D-model matches the elliptical shape and orientation angle of the main cluster. The best-fitting parameters are  listed in Table~\ref{tab:2dbetamod}; they are in very good agreement with results found by Arnaud et al.~(\cite{arnauda3921}).

\begin{table}
\begin{center}
\caption{Best-fitting results of the 2D $\beta$ model.}
\label{tab:2dbetamod}
\begin{tabular}{l|l}
\hline
Parameter& best-fitting\\
\hline
$r_{c1}$ (arcmin) & 2.80 \\ 
$r_{c2}$ (arcmin) & 2.13 \\
$\beta$ & 0.77    \\
RA  (J2000) &  $22^h50^m00^s$\\
Dec (J2000) &  $65^o$25\arcmin43\arcsec \\
PA  (rad)& 0.19\\
\hline
\end{tabular}
\end{center}
\end{table}

\begin{figure}[h]
\begin{centering}
\includegraphics[scale=0.48,angle=0,keepaspectratio]{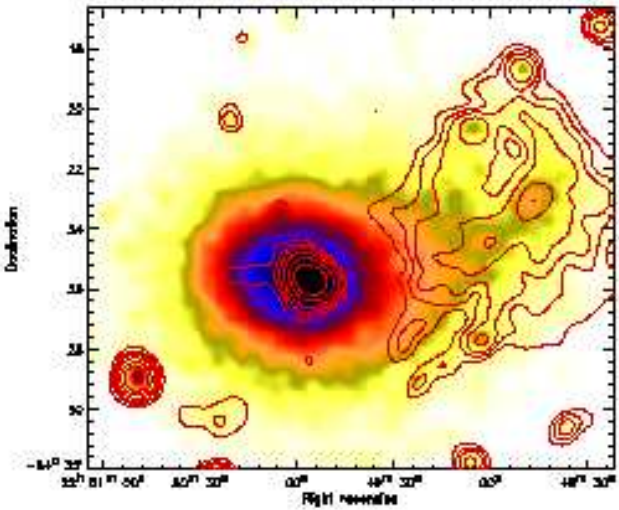}
\caption{EPIC/MOS Gaussian-filtered image with superimposed contours of the residuals after the subtraction of a 2D \betamod - see text for details. The residuals are traced at 3,5,7,10,15 and 20  $\sigma$ significance and the image is in a logarithmic scale. Epoch J2000.[\em{see the electronic version of the paper for a colour figure}]}
\label{fig:resid2D}
\end{centering}
\end{figure}

We quantified the significance of possible excess flux by subtracting the model from the image.  The residuals are shown in Fig.~\ref{fig:resid2D}. We detect extended residuals, roughly centred on the galaxy BG3, at more than 5$\sigma$ significance, supporting the previous indications of the double nature of A3921. In this framework, BG3 would be the central galaxy of a subcluster which is already highly disrupted by interaction with the main cluster, in agreement with the analysis of Arnaud et al.~(\cite{arnauda3921}). Within this extended residual structure there are two peaks, other than  BG3, one of which coincides with the position of galaxy BG2. The residuals are elongated in the direction of the main cluster to the East and display a sharp edge towards the NW. We also detect excess emission in the centre of the main cluster, reminiscent of a cooling flow. This residual emission displays a  fairly circular shape with an elongation to the east and  compression of the isophotes to the west, in the direction of the large extended residual structure. 

\subsection{1D analysis}
\label{sec:1danal}

We also performed a one-dimensional analysis by extracting a radial profile of the main cluster, binning events in circular annuli centred on the emission peak. We excluded the side of the cluster corresponding to the residuals shown in Figure \ref{fig:resid2D} (we use the main region described below and illustrated in Fig.~\ref{fig:spregions}). In view of the central excess in the 2D analysis above, it is obvious that a single \betamod\ will not be a good description of this radial profile. It was thus fitted with a double isothermal \betamod\ (the BB model, see Pratt \& Arnaud \cite{PA02}). The model was convolved with the instrument response (Ghizzardi~\cite{ghizzardi}) and binned into the same bins as the observed profile. The best-fitting values are: $r_c = 2.43^{+0.09}_{-0.10}$ arcmin, $\beta = 0.77 \pm 0.02$, $r_{\rm cut} = 1.58^{+0.10}_{-0.13}$ arcmin, and $r_{\rm c, in} = 022 \pm 0.07$ arcmin ($90$ per cent errors for one interesting parameter), which gives $\chi^2 = 223.2$ for 179 degrees of freedom. It can be seen that the external $\beta$ and $r_c$ values are in excellent agreement with the 2D analysis.

\section{Thermal structure}\label{sec:thermalstr}

In this Section, we will deal with the temperature structure of the cluster. We will first investigate the global spectral properties of the cluster. Then, after calculating a temperature map, we will extract spectra in distinct regions as an independent check of the significance of the features we find. 

In Appendix~A, we discuss the role of possible systematic effects such as variations in abundance or column density, and differences in plasma codes. Our conclusion from these tests is that the temperature structure we find is robust to these sources of systematic error.

 The $N_{\rm H}$ value used for the temperature map computation, and for any spectrum described in the following, is obtained by simultaneously fitting the MOS spectra only (between 0.3 and 10.0 keV)  with an absorbed {\sc mekal} model. This gives $N_{\rm H}$ = 2.19$_{-0.41}^{+0.42} \times 10^{20}$ cm$^{-2}$ (90 per cent errors for one significant parameter), which is $\sim$ 25 per cent lower than the Galactic value given by Dickey \& Lockman (1990), but in excellent agreement with the PSPC result of $N_{\rm H} = (2.2\pm0.2) \times 10^{20}$ cm$^{-2}$ (Arnaud et al.~\cite{arnauda3921}). Possible systematics related to $N_{\rm H}$ variations are described in Appendix~A.

\subsection{Global spectrum}\label{sec:spglob}

In this Section we are interested in evaluating the global properties of the cluster, for comparison with previous works, and to quantify the effect of the merger on them.

 The spectral analysis discussed in this Sect. and the following is performed by extracting spectra from the weighted source and background (blank-sky) event lists. Spectra were extracted in each region, and the background spectrum was normalised and subtracted from the source spectrum. The spectra of the three cameras were fitted simultaneously in the (0.3/0.5 - 10.0) keV range (MOS/pn).  

We extracted a spectrum in a large circle of radius 11\arcmin, corresponding to $\sim 0.65~r_{200}$\footnote{Here the virial radius is defined as r$_{200} = h(z)^{-1} \times 3.69 \times (T/10~{\rm keV})^{1/2} h^{-1}_{50}$ Mpc, where $h(z) = (\Omega_m \times (1+z)^3 + \Lambda)^{1/2}$ and the relation is scaled from Evrard, Metzner \& Navarro~ (\cite{emn96}).}. We will refer to this spectrum as the {\em global} spectrum. 

After considerations about possible systematics errors due to different plasma codes and/or specific element variations (tests are described in detail  in Appendix~A), we fitted the global spectrum with a single and a two-temperature {\sc mekal} model. Results are listed in Table \ref{tab:spregmodels}. The single-temperature model gives a reduced  $\chi^2$= 1.36,  clearly not satisfactory. Adopting this model, the global temperature, k$T = 4.34\pm0.10$ keV (90 pe cent errors), is in agreement with the GINGA (k$T = 4.9\pm0.6$) and ROSAT (k$T = 4.1\pm1.0$) results of Arnaud~\etal~(\cite{arnauda3921}), but is significantly lower than the ASCA results (k$T = 5.73\pm0.24$; White 2000). 

\begin{figure}
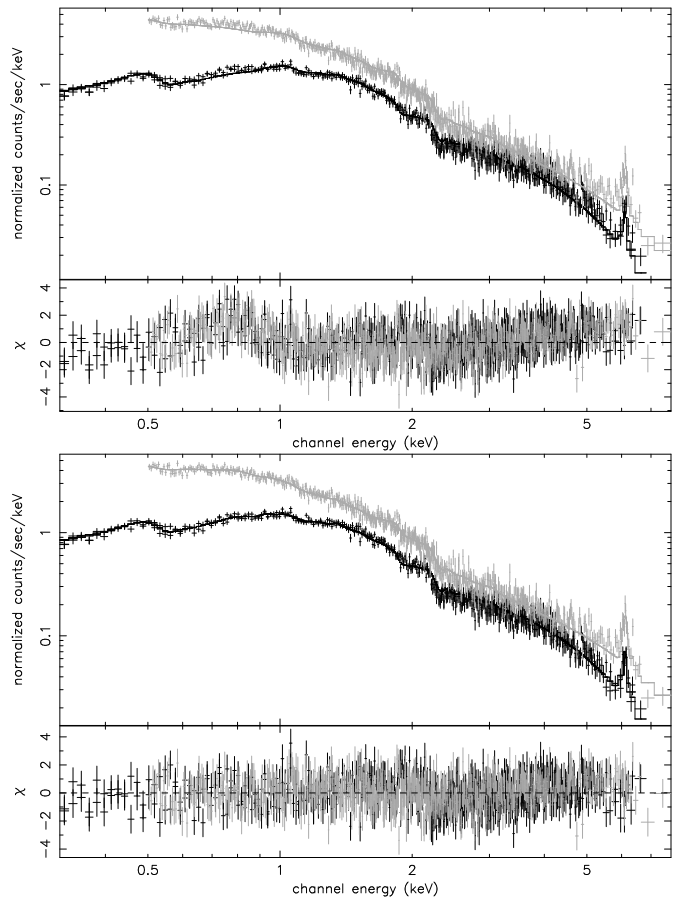

\centering
\includegraphics[scale=0.4,angle=-90,keepaspectratio,width=\columnwidth]{f3a.ps}
\includegraphics[scale=0.4,angle=-90,keepaspectratio,width=\columnwidth]{f3b.ps}
\caption{Global spectrum integrated in a circular region of radius 11\farcm2, corresponding to $~\sim 0.65~r_{200}$. The pn is in grey the MOS in black. {\bf Top:} The folded model is the single-temperature {\sc mekal}. One can observe the large residuals at $\sim$ 0.8 keV. {\bf Bottom:} The folded model is the two-temperature {\sc mekal}. The spectra are binned to have a $S/N$ ratio of 5$\sigma$ above the background.}
\label{fig:spglobfit}
\end{figure}

With the single temperature fit to the global spectrum, we observe large residuals corresponding to the Fe L multiplet (see the top panel of Fig. \ref{fig:spglobfit}), reflecting the existence of multi-temperature gas as observed in the temperature map above. The largest residuals are centred at $\sim 0.8$ keV. If the Fe-L-shell lines are used as a thermometer (Belsole et al. 2001, B\"ohringer et al. 2002), we might expect to observe at least another component at $\sim$ 0.6 keV. We thus fitted the spectrum with two-temperature {\sc mekal} model, obtaining  significantly better fit (reduced $\chi^2$=1.1). We observe that both the dominant temperature and the chemical abundances are higher in this case (but see Appendix~A). The two-temperature fit is shown in the bottom panel of Fig.~\ref{fig:spglobfit}.  The low temperature (k$T$ = 0.65 keV) and luminosity ($\Lx = (1.7\pm0.2)\times 10^{43}$ \es, more than an order of magnitude lower than the first component)  of the low-temperature thermal component may originate from the central galaxy. To test this hypothesis, we fitted the central bin  of the temperature profile (see Sect. \ref{sec:tprof}), e.g. the region within 30 arcsec, with one and two  {\sc mekal} models. The spectrum is well fitted with a 1-temperature model of k$T=5.44$ keV (not requiring a second component), indicating that any galaxy component is negligible with respect to the cluster emission.  The low-temperature component more likely arises from the fact that we are fitting a multi-temperature plasma with a two-temperature model (see also Appendix).

White (2000) detected a cooling flow with a large mass deposition rate, and  we see a residual in the core in the 2D $\beta$-model fit. However, our temperature map (and profile) clearly shows that there is no cool gas in the core.

\subsection{Temperature map}\label{sec:tmap}

\begin{figure*}[!]
\centering
\includegraphics[scale=1.0,angle=0,keepaspectratio]{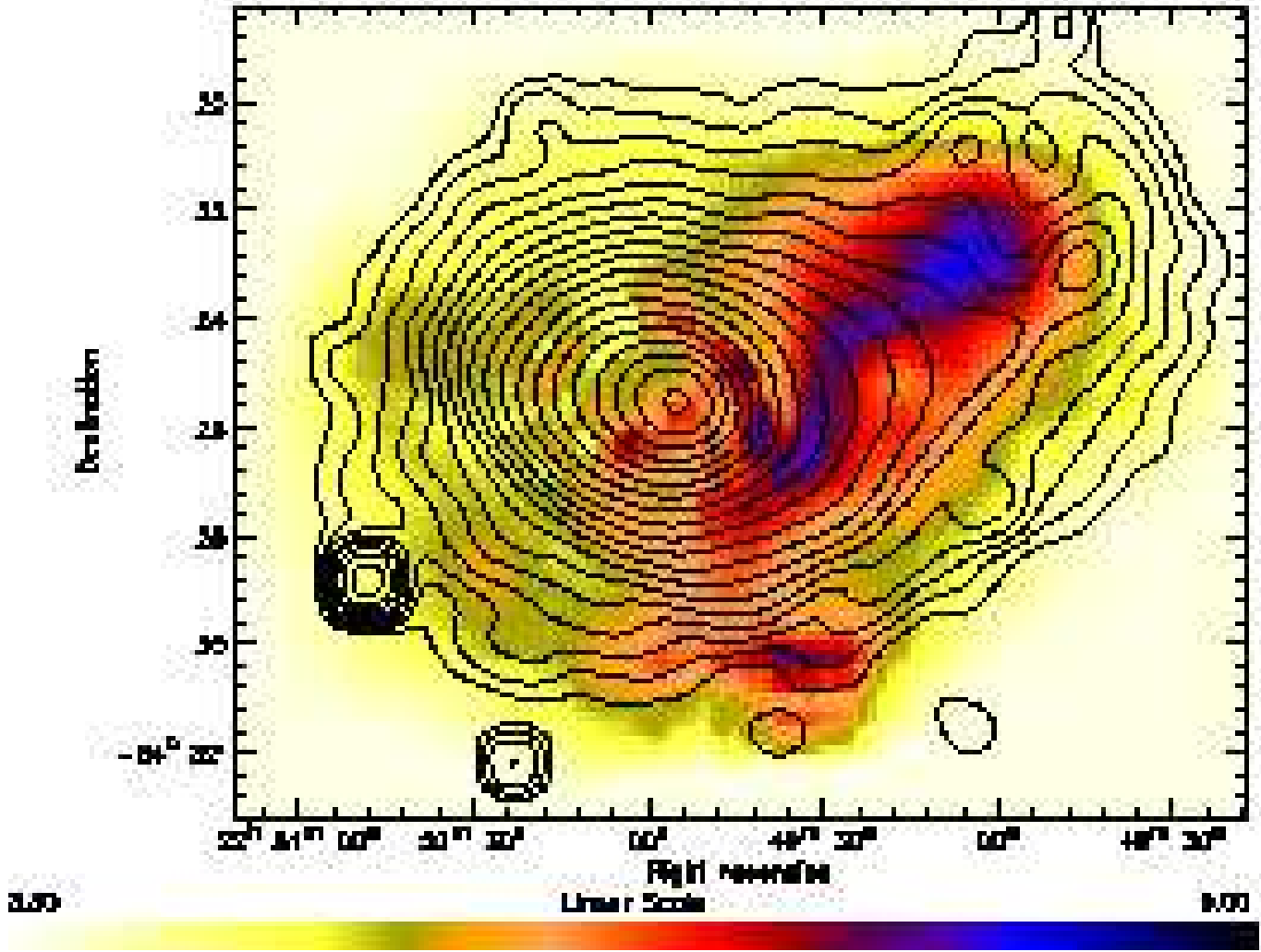}
\caption{Temperature map obtained by applying the wavelet algorithm described in Bourdin et al. (\cite{bourdin04}). The contours are obtained from the MOS adaptively smoothed image in the energy band 0.3-10 keV, and are logarithmically spaced, as in Fig. \ref{fig:fig1}. The first contour corresponds to 0.6 (MOS1+MOS2) counts/pixel. Image coordinates epoch is J2000. [\em{see the electronic version of the paper for a colour figure}].}
\label{fig:tmap}
\end{figure*}

The temperature map was obtained by applying the multi-scale spectro-imaging algorithm described in Bourdin et al. (\cite{bourdin04}). Details of the application of this technique to \xmm\ data can be found in Belsole et al. (\cite{a1750}). Although we did not include the pn camera because of technical considerations,  in the case of A3921 we have more counts than we had for A1750 and the temperature structure is mapped in finer detail. To enhance the stability of our results we impose a (conservative) lower limit of 1000 photons per pixel at each spatial scale. Both the particle and X-ray backgrounds are modelled in this approach, allowing us to achieve typical errors of $\sim 0.1$ keV per pixel (see Bourdin et al. \cite{bourdin04}, for details). The fit in each pixel and spatial scale was performed with the metallicity fixed to the best-fitting single-temperature {\sc mekal} fit to the global spectrum, and the column density fixed to the best-fitting value derived from the MOS spectral fit to the {\em main} spectrum ($N_{\rm H} = 2.19 \times 10^{20}$ cm$^{-2}$; see Appendix~A for details).

The temperature map obtained from this wavelet analysis is shown in Fig.~\ref{fig:tmap}. The overlaid contours are the same as in Fig.~\ref{fig:fig1b}. The overall appearance of the temperature structure is strongly asymmetric. We observe an extended hot region which appears related to the extended residuals seen in Fig.~\ref{fig:resid2D}. The core of the main cluster is also hotter than the surroundings, at least to its south-western side. The hottest region of the map is oriented nearly parallel to the line joining the centre of the main cluster and a point between the two dominant galaxies to the west, and indeed terminates in a region between BG2 and BG3. In the direction of the main cluster, at 1.5\arcmin~ ($\sim 150 h_{75}^{-1}$ kpc) from the centre towards the west, the hot emission is more elliptically shaped and follows the X-ray emission isophotes. The eastern side of the main cluster is fairly isothermal, with an average temperature of 4.9 keV. The overall temperature structure is difficult to understand if the main cluster and the secondary structure are just starting to collide. 

We also detect a hot region to the south of the core, at around 5.3\arcmin~ (RA = $22^h49^m35.8^s$, Dec = $-64^o30\arcmin25\farcs35$ (J2000)). We did not find any correspondence between this hot spot and any X-ray emission peak. We notice that the emissivity contours are distorted at this location. In an attempt to quantify these emissivity distortions, we generated a radial profile where we azimuthally integrated counts within a sector (of angle 60$^{\circ}$) centred at the cluster centre  and including the hot region. We did not find significant deviation from a $\beta$-model. We also compared with the deep optical image (\cite{f04}), but there is no optical counterpart which could explain a hot emission at this location. Thus everything seems to indicate that this hot emission is related to the merger event.

\subsection{Region analysis}\label{sec:discrete}

\begin{figure}
\centering
\includegraphics[scale=0.45,angle=-90,keepaspectratio,width=\columnwidth]{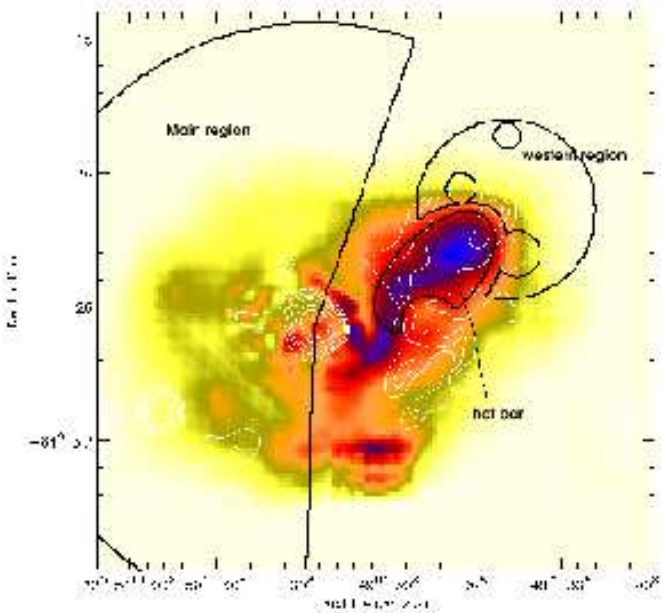}
\caption{Figure illustrating discrete regions from which spectra were extracted. The regions were defined on the basis of the residual morphological and temperature structure [\em{see the electronic version of the paper for a colour figure}].}
\label{fig:spregions}
\end{figure}

We now turn to an assessment of the significance of the temperature structure found in the temperature map. In the following analysis, we will concentrate on spectral fits to discrete regions with {\sc mekal} models. Except where noted, the column density is fixed to the best-fitting value derived from the MOS spectral fit to the {\em main} spectrum ($N_{\rm H} = 2.19 \times 10^{20} cm^{-2}$). The individual regions are illustrated in Fig. \ref{fig:spregions}.

\subsubsection{Discrete region 1: the main cluster}

We extracted a spectrum for the main cluster only by selecting a sector between PA = $71^{\circ}$ and $268^{\circ}$ (East to North, illustrated in Fig. \ref{fig:spregions}), e.g. excluding the western extension and hot emission bar. We will refer to this spectrum as the {\em main} spectrum. We fitted the {\em main} spectrum with single and double {\sc mekal} models, for which the results are detailed in Table ~\ref{tab:spregmodels}. 
The {\em main} region has a temperature which is slightly lower than the {\em global} temperature, a direct consequence of excluding the hot bar (see Fig. \ref{fig:tmap}). This temperature is also lower than the average temperature on the eastern side of the temperature map. Part of this discrepancy can be explained by the fact that we did not use the pn camera when deriving the temperature map. Once again the two-temperature model is a better representation of the data ($F$-test probability $>99.9$ per cent), suggesting the existence of multi-phase gas even on the side of the cluster which looks the most dynamically relaxed. The ratio between the bolometric X-ray luminosity found in the global and main regions is $L_X^{global}/L_X^{main}$ = 2.0 which is comparable to the ratio of the two areas of integration.  

\begin{table*}
\begin{center}
\caption{Best-fitting results of discrete regions.}
\label{tab:spregmodels}
\begin{tabular}{lcccccccr}
\hline
region & T1   & Z1 	& Norm1 			 &T2 	& Z2 	& Norm2 & $L_{\rm X}^{bol}$ & $\chi^2$/d.o.f. \\
       & (keV)& solar	& ($^*$) &(keV) &solar  &($^*$)	& (10$^{44}$ erg s$^{-1}$) &  \\
\hline
global &  4.38$^{+0.07}_{-0.07}$ &   0.26$^{+0.03}_{-0.03}$ & 1.86$_{-0.02}^{+0.02}\times10^{-2}$ &--- &--- & ---& 5.24$^{+0.10}_{-0.10}$& 1551.1/1140  \\
global &  5.18$_{-0.17}^{+0.18}$ & 0.39$_{-0.04}^{+0.04}$ &1.70$^{+0.04}_{-0.03}\times10^{-2}$ & 0.65$_{-0.04}^{+0.04}$&0.17$_{-0.06}^{+0.15}$ & 0.14$_{-0.06}^{+0.06}\times10^{-2}$ & 5.54$^{+0.10}_{-0.10}$&1256.1/1136\\
main & 4.16$^{+0.10}_{-0.10}$ & 0.25$^{+0.04}_{-0.04}$ & 0.97$^{+0.01}_{-0.01}\times10^{-2}$ & --- & --- & --- & 2.65$^{+0.10}_{-0.10}$&1190.2/898 \\
main & 4.90$_{-0.25}^{+0.26}$ & 0.39$_{-0.06}^{+0.07}$ &0.87$^{+0.02}_{-0.02}\times10^{-2}$ & 0.68$_{-0.06}^{+0.05}$ &0.14$_{-0.06}^{+0.17}$&0.10$_{-0.05}^{+0.05}\times10^{-2}$  & 3.80$^{+0.10}_{-0.10}$ & 1063.1/894\\
hot bar & 6.30$^{+0.48}_{-0.45}$ & 0.28$^{+0.13}_{-0.13}$ & 1.20$^{+0.04}_{-0.04}\times10^{-3}$ & --- & --- & --- &0.40$^{+0.02}_{-0.02}$ & 708.2/707 \\
hot bar & 7.25$_{-0.62}^{+0.79}$ & 0.36$_{-0.14}^{+0.12}$ &1.14$^{+0.03}_{-0.04}\times10^{-3}$ & 0.64$_{-0.17}^{+0.22}$ & --- &4.11$_{-2.05}^{+2.96}\times 10^{-5}$  & 0.42$^{+0.05}_{-0.04}$& 687.0/704\\
western & 3.60$^{+0.39}_{-0.29}$ & 0.32$^{+0.19}_{-0.16}$ & 7.68$^{+0.47}_{-0.46}\times10^{-4}$ & --- & --- & --- &0.20$^{+0.02}_{-0.02}$ & 419.9/384 \\
western & 4.59$_{-0.48}^{+0.61}$ & 0.73$_{-0.21}^{+0.28}$ &6.35$^{+0.63}_{-0.34}\times10^{-4}$ & 0.67$_{-0.01}^{+0.09}$ & --- &3.25$_{-1.21}^{+1.70}\times10^{-5}$  & 0.22$^{+0.03}_{-0.03}$& 387.6/381\\

\hline
\hline
\end{tabular}
\begin{minipage}{17.5 cm}
Errors are quoted at 90 per cent confidence for one interesting parameter. For the hot bar and the western regions the abundances of the two thermal components were tied.$^*$The normalisation is in units of $10^{-14}/ (4\pi (D_A\times(1+z))^2$) cm$^{-5}$, where D$_A$ is the angular size distance to the source (cm), as defined in the {\sc mekal} model, in XSPEC. 
\end{minipage}
\end{center}
\end{table*}

\subsubsection{Discrete region 2: the hot bar}

We next extracted the spectrum of the {\em hot bar} observed in the temperature map. The region is an ellipse of semi-minor axis 98\arcsec, semi-major axis  174\arcsec, and position angle 45$^{\circ}$, centred on RA = $22^h49^m14^s813$, Dec = $-64^{\circ}23\farcm49\farcs51$ (see Fig. \ref{fig:spregions}). Results are listed in Table \ref{tab:spregmodels}. As expected, the temperature of the {\em hot bar} is significantly higher than the {\em global} and {\em main} cluster regions. Possible systematics can arise because of absorption variations (see Appendix~A). To test this effect, we left the $N_{\rm H}$ free to vary, obtaining a best-fitting $N_{\rm H}$ lower than (but consistent with) the average value and an even higher temperature (although with larger errors). Thus if there are any systematics related to absorption variations, the approach of fixing the $N_{\rm H}$ for the derivation of the temperature map is somewhat conservative.

When a second thermal component is added to the model, the dominant component reaches  a temperature of 7.25 keV. The $F$-test indicates that the two-temperature model is a better representation of the data, with a probability of $3.14 \times 10^{-3}$ that the second component is detected by chance. Figure \ref{fig:tmap} has already graphically illustrated the multi-temperature nature of the hot bar.

\subsubsection{Discrete region 3: the western region}

In order to further verify the asymmetric temperature distribution represented by the hot bar, we extracted a spectrum in the region to the west, beyond the hot bar, as shown in Fig. \ref{fig:spregions}. We excluded the X-ray emission coincident with the two brightest galaxies BG2 and BG3. 

This region is slightly cooler than the {\em main} region (but consistent within the errors). Once again, we observe that a two-temperature thermal model is a better representation of the data than a single temperature model ($F$-test probability $>$99.9 per cent).  We also observe an increase in relative abundances, which might suggest chemical contribution from a more metal-rich object at the location of the residual structure.

\subsection{Temperature profile}\label{sec:tprof}

The gas density and temperature profiles are key observational quantities for the derivation of many other cluster characteristics. In particular, these quantities are essential for the derivation of mass and entropy profiles. In this Section, we will calculate the temperature profile of A3921, and assess possible systematics. Most analyses of more relaxed clusters assume single temperature models, and so we ignore multi-temperature effects in the present analysis. 

\begin{figure}
\centering
\includegraphics[scale=0.45,angle=0,width=\columnwidth,keepaspectratio]{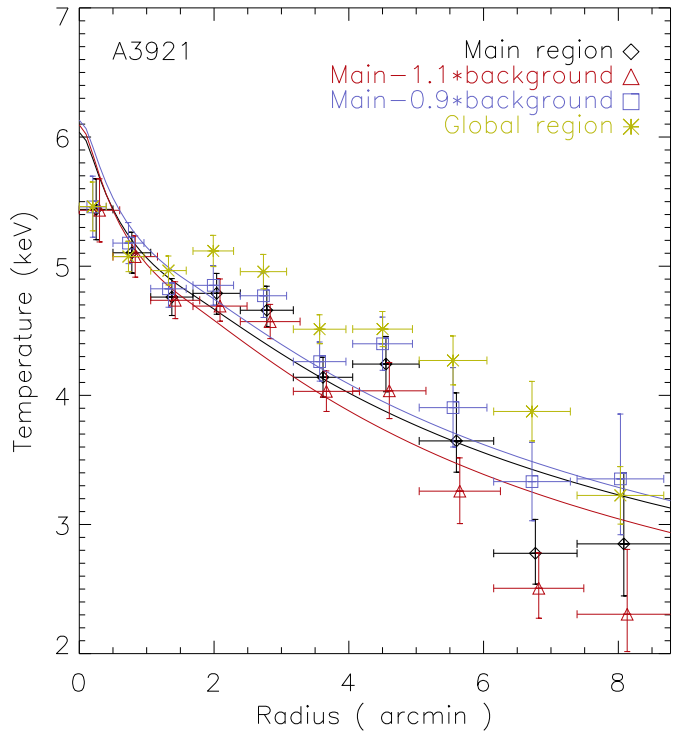}
\caption{Projected radial temperature profile of the main and global regions. The lines overplotted on the main region profiles are the best-fitting polytropic models. The values of $T_0$ and $\gamma$ for all of these models agree within their $1 \sigma$ errors (see text for details).}\label{fig:tprof}

\end{figure}

We first obtained a temperature profile in the {\em main} region, 
(Fig. \ref{fig:spregions}). The centre of the profile was fixed to the peak of emission of the main cluster. The radial temperature bins were defined by requiring a signal-to-noise of 5$\sigma$ in the 2.5-6.5 keV radial profile, with a lower limit of $30\arcsec$ to minimise Point Spread Function (PSF) effects. (The annuli thus have widths greater than or equal to the diameter enclosing 70 per cent of the energy for the on-axis PSF). We were able to extract spectra in 10 annuli, out to a distance which corresponds to $\sim 0.5~r_{200}$. We fitted each spectrum with a single {\sc mekal} model where the column density was fixed to the value used for the temperature map and the discrete regions ($N_{\rm H} = 2.19 \times 10^{20}$ cm$^{-2}$). Chemical abundances were left as free parameters out to annulus 8 and then fixed to the fitted value obtained for annulus 5.\footnote{When the abundances were fixed to 0.25, the fitted temperature were statistically consistent with the values obtained when the abundances were left free.} The three camera spectra were fitted simultaneously, ignoring channels below 0.3 for MOS and below 0.5 for pn.
We found excellent $\chi^2$ for all our bins, indicating that a second thermal component is not required if the spectrum extraction region is sufficiently small. The resulting projected temperature profile is shown in Fig.~\ref{fig:tprof}. 

The quiescent background level of \xmm\ can vary by $\pm 10$ per cent. We thus calculated temperature profiles with the background at plus and minus 10 per cent of our adopted normalisation. The effect is negligible, both of these profiles are consistent with the nominal profile at $1 \sigma$, as can be seen in Fig.~\ref{fig:tprof}. For easier comparison with previous work, we also extracted a temperature profile from annular regions encompassing the entire cluster (the {\em global} region). This is also shown in Fig.~\ref{fig:tprof}.

There is no sign of a cool core region, and all the profiles drop almost monotonically to about $50$ per cent of the central value at the detection limit. The {\em global} region profile is slightly hotter and flatter than that of the {\em main} region, as expected since this region includes the emission from the hot bar. However, the temperature still declines with radius. This is different from the flat temperature profile found by White (2000) with ASCA. We have fitted a polytropic model (e.g., Markevitch et al. \cite{marketal98}), $T = T_0 \times \rho^{\gamma -1}$, to the {\em main} region profiles. The gas density profile was described by the BB model discussed in Sect.~\ref{sec:1danal}, with all parameters fixed. We find that the nominal main region profile has $T_0 = 6.07\pm0.03$ keV and polytropic index $\gamma=1.16\pm0.01$. The values of $T_0$ and $\gamma$ for the {\em main} region profiles with background at plus/minus 10 percent of nominal agree within their $1 \sigma$ errors. 

\subsection{Luminosity}

Numerical simulations (Ritchie \& Thomas \cite{RT02}; Randall, Sarazin \& Ricker~ \cite{rsr02}) suggest that the effect of the merger is to boost the luminosity and the temperature for a short time, and this can have an obvious effect on scaling relations which involve these quantities. Arnaud \& Evrard (\cite{AE99}) found  $\Lx \propto T^{2.88}$ using a sample of clusters with weak or absent cooling flow signatures. Their relation has an intrinsic dispersion of 0.13 in log $\Lx$ at a given temperature. 
Taking the simple one-temperature fit to the global and main regions in our analysis and scaling the $\Lx - \Tx$ relation of Arnaud \& Evrard (\cite{AE99}) for our cosmology, we  find that A3921 is a factor of two too luminous for its temperature. We further note that this cluster is an outlier in their analysis (see their Fig. 1), but less significantly so because of the larger errors on their temperature measurement.

\section{Mass analysis}

\begin{figure}
\centering
\includegraphics[scale=0.45,angle=0,width=\columnwidth,keepaspectratio]{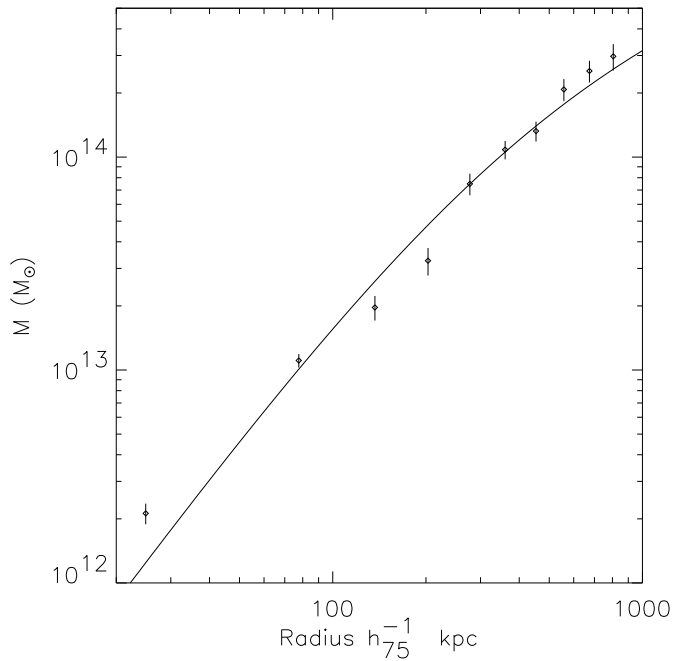}
\caption{Integrated total mass profile (points with errors), using data from the unperturbed part of the cluster. The solid line is the best-fitting NFW profile ($\chi^2_{\nu} \sim 4.5$).}\label{fig:mprof}
\end{figure}

Combining a gas density (Sect.~\ref{sec:1danal}) and a temperature (Sect.~\ref{sec:tprof}) profile, we can derive a total gravitational mass profile under the assumptions of hydrostatic equilibrium and spherical symmetry.  Given the observed ellipticity and substructure (Sect.~\ref{sec:2Dbeta}) and the strong merger signatures (Sect.~\ref{sec:tmap}), it is very likely that the cluster is {\it not\/} in hydrostatic equilibrium. We have thus tried to minimise these effects by excluding the regions most strongly affected by the merger event. This approach has often been used for clusters with substructure (e.g., Henry, Briel \& Nulsen~\cite{hbn93}), and the approach has been shown to give reasonable results when applied to simulated clusters (Schindler~\cite{ss96}).

We thus use the {\em main} region to derive the mass profile. The mass was calculated at each radius of the temperature profile with the Monte Carlo method described in Pratt \& Arnaud~(\cite{PA03}), which takes as input the parametric model for the gas density and the measured temperature profile with errors. A random temperature is calculated at each radius of the measured temperature profile, assuming a Gaussian distribution with sigma equal to the $1\sigma$ error, and a cublic spline interpolation is used to compute the derivative. Only profiles corresponding to a monotonically increasing mass gradient are kept: 1000 such profiles were calculated. The final gravitational mass profile is shown in Fig.~\ref{fig:mprof}. 

For this cluster at least, differences in temperature profiles have a negligible effect on the resulting mass profile. Mass profiles calculated using all of the temperature profiles detailed above in Sect.~\ref{sec:tprof} (including that of the {\em global} region which contains all the disturbed temperature structure) all agree within the $1\sigma$ errors.

We have tried to fit the mass profile with several different functional forms for the dark matter density. The Moore et al.~(\cite{mqgsl}) and King approximation to an isothermal sphere models are unconstrained. An NFW model (shown in Fig.~\ref{fig:mprof}) yields $c=4.3$ and $r_s = 337 h_{75}^{-1}$ kpc, but has a reduced $\chi^2_{\nu} = 4.5$. Thus either i) the density profile of this cluster is not well described by the NFW model, or ii) even in the relaxed-looking {\em main} region, the gas is not in hydrostatic equilibrium. In any case, the poor fit argues in favour of the cluster being dynamically perturbed.

It is important to understand the effect of mergers on mass calculations. The NFW fit is particularly poor in the inner regions ($r \lesssim 150 {\rm kpc}$), but beyond this radius it is a rather good description of the X-ray mass profile. This suggests that it might be interesting to compare the total optical and X-ray masses within the fiducial optical radius of $0.34 h_{75}^{-1}$ Mpc (\cite{f04}). The NFW fit yields $9.8 \times 10^{13} h_{75}^{-1} M_{\odot}$ interior to this radius, while a linear extrapolation of the X-ray mass profile in the log $r$-log $M$ plane gives $(9.9 \pm 1.2) \times 10^{13} h_{75}^{-1} M_{\odot}$. The optically-derived mass of the main cluster within this radius is $2.1^{+1.3}_{-0.9} \times 10^{14} h_{75}^{-1} M_{\odot}$. Thus at this radius the different mass estimators agree within their respective errors.

\section{The centre of the main cluster}\label{sec:centre}

In Sect. \ref{sec:2Dbeta} we found that the centre of the cluster shows an excess of emission over a $\beta$-model, and that this region also has a relatively  high temperature (5.4 keV), as shown by the temperature profile and map, which also suggest that this central region is in a multi-phase state. In principle, while the high temperature of the first bin  in the temperature profile might be produced by an active galactic nucleus (AGN) in the central galaxy BG1, we can find no evidence for a second component in the spectrum from this bin. In addition, the large extent of the spatial residuals (2 arcmin $\sim 130$ kpc) make it unlikely that what we observe could be produced by BG1 alone. We thus conclude that  the high central temperature is likely a consequence of the merging event.

\begin{figure}
\begin{center}
\includegraphics[scale=0.35,angle=-90,keepaspectratio]{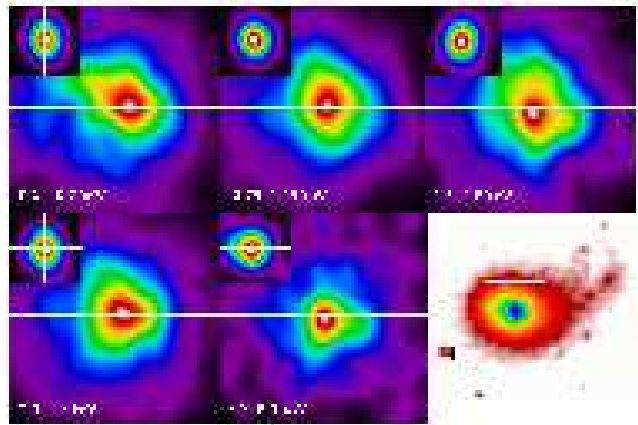}
\caption{Images of the central 200 kpc of A3921. All images are smoothed with a Gaussian filter of $\sigma = 11 \arcsec$. From top left to bottom centre the energy bands are: 0.2 - 0.7 keV; 0.75 - 1.25 keV; 1.3 -1.8 keV; 2.0 -4.5 keV and 4.7-8.0 keV. The insert in each figure shows the corresponding point source image. The bottom right image shows the regions used for the image extraction: in black for the point source and in white for A3921.[\em{see the electronic version of the paper for a colour figure}]}
\end{center}
\label{fig:core}
\end{figure}

We looked at the central 450 kpc (a $4\farcm5\times4\farcm5$ square region) in more detail by extracting images in 5 different energy bands. This allows us to compare simultaneously the spectral and spatial structures. In Fig. \ref{fig:core} we show from the top left to the bottom centre the images in the (0.2-0.7) keV, (0.75-1.25) keV, (1.3-1.8) keV, (2.0-4.5) keV and (4.7-8.0) keV bands obtained with the MOS1 camera. The images have been smoothed with a Gaussian filter of $\sigma$ = 11\arcsec. The cross is centred on the emission peak in the (0.2-0.7) keV image. In order to check for PSF effects, we also  produced images in the same energy bands for a bright point source approximately 9\arcmin~ off-axis in the same field (see Fig. \ref{fig:core}, bottom right) - note that the X-ray peak of A3921 is 4\arcmin~ off-axis. The point source image is shown in the insert for each energy band. 

Despite the fact that the point source is further off-axis than the X-ray peak of A3921, we note that its centroid position does not change with energy (except perhaps for the hardest energy band, where we measure a small shift of 3\farcs3 toward the East). The centroid of the A3921 emission peak changes significantly with energy. The largest shift (excluding the hardest energy band) is 15\arcsec~ for the (1.3-1.8) keV image. We also observe isophotal orientation variation. 

The lowest-energy image is the most sensitive to small-scale changes in the column density, and the lower than Galactic $N_{\rm H}$ we have found in the global spectral fit supports this hypothesis. This difference in  $N_{\rm H}$ values is a possible explanation for the shifted centre position with respect to other energy bands. The (0.75-1.25) keV image covers to the Fe L multiplet. The centroid shift and the North-South elongation of the emission in this energy band is possibly connected to abundance variations, or, more likely, temperature variations.
The shift in the hardest energy band is partially due to the PSF, as observed for the point source (in the insert). However, PSF effects cannot explain the whole shift. For the other two energy intervals, we cannot find an obvious physical explanation for the centroid shift, apart from perturbations due to the merger event. The {\em Chandra} observation of A3921, to be performed shortly, will shed more light on this issue.

\section{The bright galaxies BG2 and BG3 in X-ray} \label{sec:galaxies}

We have also examined the X-ray properties of the bright galaxies BG2 and BG3. We extracted the spectrum of each galaxy in circular regions of radius of 20\arcsec\ and 16\arcsec\ for BG2 and BG3 respectively. As a background we selected an annular region of inner radius 33\arcsec\ and outer radius 53\arcsec\, which takes into account the cluster emission at the location of the galaxies.

There are 300 net MOS+pn counts from  BG2 between 0.3 and 5.0 keV, severely limiting the constraints we can put on the spectral parameters. We fitted simultaneously the three camera spectra with an absorbed {\sc mekal} model. We fixed the galactic absorption to the global X-ray value (2.19$\times 10^{20}$ cm$^{-2}$),  and let the temperature and abundance vary. The best-fitting gives k$T = 1.10_{-0.18}^{+0.37}$ and a low metallicity of 0.05$(<0.14)$ solar, with a $\chi^2 = 12.5/21$ d.o.f (errors are quoted at 1$\sigma$ confidence). If we fix the metallicity to be 0.3, the $\chi^2$ rises to 15.6  for 22  d.o.f, and the best-fitting temperature is higher (1.34$_{-0.23}^{+0.64}$ keV).  

We are also interested if the X-ray emission has a non-thermal origin and thus we also fitted the spectrum with a power-law model. We obtain a photon index $\gamma = 2.26^{+0.36}_{-1.26}$, consistent with X-ray results of active nuclei (e.g. Brinkmann, Papadakis \& Ferrero \cite{BPF04}). The $\chi^2$ is 11.1/19 d.o.f. In this case, the thermal component best-fitting parameters are k$T$ = 0.9($\pm0.4)$ keV and $Z=0.7(>0.13$). The thermal and non-thermal component have similar normalisations.

The thermal and non-thermal models fit the data equally well, but it has to be admitted that the data are not very constraining. However, the non-thermal model implies that BG2 harbours an active nucleus. A3921 was observed with the Australia Telescope Compact Array (ATCA) at 2.5 GHz and 1.5 GHz. The 2.5 GHz radio map (M. Hardcastle, private communication) shows faint radio emission in the centre of the optical counterpart of BG2. No radio emission is detected in the lower frequency observation (1.5 GHz), suggesting that the source may indeed be an AGN. We conclude that the X-ray emission from BG2 may be non-thermal in origin, more likely coming from a central active nucleus. The weakness of the X-ray and radio emission could be a consequence of the merging event. If the gas surrounding the core of BG2 is stripped by the merger, the active nucleus would be starved of fuel.

The X-ray spectrum of BG3, on the other hand, looks more thermal. The statistical quality here is slightly better, with 406  MOS+pn photons in the (0.3-3.0) keV energy range. We fitted simultaneously the three camera spectra with an absorbed {\sc mekal} model, the $N_{\rm H}$ fixed as above. The best-fitting temperature is k$T = 1.07^{+0.21}_{-0.20}$ keV (1$\sigma$ confidence), and the abundances $Z=0.11^{+0.18}_{-0.07}$ solar, for a $\chi^2= 25.5/26$ d.o.f. With this model we calculate a bolometric luminosity of 6.0$\times10^{41}$ erg s$^{-1}$. Since contamination from the cluster may not have been completely accounted for by the background region we have chosen, we also fitted the spectrum with a two-temperature model where the high temperature was fixed to the cluster temperature at the position of BG3. We found that this model is a slightly (not significant) better representation of the data ($\chi^2= 24.8/25$). The temperature and abundances are in agreement with the 1-temperature model, k$T=0.90^{+0.33}_{-0.17}$ keV, $Z=0.48^{+0.26}_{-0.24}$. 

The relatively high  X-ray luminosity suggests that BG3 belongs to the category of X-ray bright galaxies (e.g. Forman, Jones \&  Tucker \cite{formanetal85}; Trinchieri, Fabbiano, \& Canizares \cite{trinchierietal86}), for which the diffuse X-ray emission from the IGM dominates over the emission of a population of non-resolved sources (mainly low mass X-Ray binaries; e.g., Sivakoff, Sarazin, Irwin \cite{sivakoff03}). However, the emission around BG3 looks more extended than that of a normal early-type galaxy. With our data we cannot exclude the possibility that the relatively high X-ray temperature is due to the emission from a small galaxy group with BG3 at its centre, and which presumably has not yet collided with the main cluster or subcluster. It is difficult with the present data to separate any possible group emission from the extended emission of the X-ray residuals (see Fig. \ref{fig:resid2D}). However \cite{f04} do not find other bright galaxies around BG3 in the projected redshift space. Deeper optical observations in the region round BG3, together with the upcoming {\em Chandra} observation, will help shed light on the nature of BG3.

\section{Discussion}
\subsection{A new picture}
The morphological analysis of the \xmm\ observation (Sect. \ref{sec:2Dbeta}) has shown unequivocally that there are two peaks in the X-ray emission from the system. The bulk of the X-ray emission is centred on the cD galaxy BG1. The residuals above a simple 2D \betamod\ fit to the main cluster are centred on BG3, which is clearly visible as an extended source even in the raw image (Fig. \ref{fig:fig1a}). 

The temperature analysis (Sect.~\ref{sec:tmap}) has shown a  complicated temperature structure to the West of the main cluster, coincident with the residuals from the morphological analysis. In particular, we detect  a ``bar'' of significantly higher temperature gas (k$T$ from 7 to 8  keV) which describes an angle which is approximately parallel to the line joining BG1 and BG2. If we draw a line in the SE-NW direction, passing through BG1, we see that the temperature structure above the line is far more uniform than that below the line. Our analysis in discrete regions combined with the stable temperature fit after considering several possible systematic effects firmly demonstrates that the cluster temperature distribution is strongly asymmetric, with the temperature of the hot bar being more than 2 keV higher than the less perturbed side of the cluster.

The centre of the main cluster shows strong evidence for having been perturbed, both in shifts of the X-ray centroid with energy (Sect.~\ref{sec:centre}), and in the offset between the position of BG1 with that of the overall X-ray centroid. We note that the offset is in the same SE-NW direction as the hot bar. Furthermore, the monotonically-rising temperature profile towards the centre is neither in agreement with other \xmm\ observations of relaxed clusters, nor with the concept of a cooling core region, a natural conclusion from the surface brightness analysis (Sect.~\ref{sec:2Dbeta} and \ref{sec:1danal}) alone. Moreover, the central regions of the mass profile (derived from the relaxed-looking {\em main} region) are not well fitted by the NFW profile, which is a reasonable representation of the mass distribution at larger radii.

The optical observations of \cite{f04}  have given firm evidence for two peaks in the galaxy density distribution, one centred on BG1 (main component, A) and the other centred around BG2 (component B),  the two components being at the same mean redshift.

If we did not know the galaxy redshifts, it would be tempting to conclude that the residual emission we see is a secondary diffuse X-ray component with BG3 at the centre, as was earlier suggested by Arnaud et al. (\cite{arnauda3921}). BG3 lies at a significantly different redshift from that of the two main galaxy peaks. From the current spectroscopic results  of \cite{f04} there is no evidence at the moment for a concentration of galaxies around BG3 in the redshift space. Other galaxies are measured to have the same $cz$ as BG3, but they are located more in the main cluster central region (see \cite{f04} for details). However, the X-ray temperature of BG3 is rather high (Sect.~\ref{sec:galaxies}), which could mean that it is associated with a group. Deeper optical observations should resolve the issue.

\subsection{Tentative interpretation}

Previous detailed work on this cluster (Arnaud et al.~\cite{arnauda3921}) suggested that the system is in an early merger phase, with the cluster B on its first infall towards the main cluster. With the present \xmm\ observation and in conjunction with results presented in \cite{f04}, we have several pieces of evidence that this interpretation is incorrect.

The existence of the hot bar and particularly its orientation is the strongest evidence that we are not observing an early stage of the merger event. If we were, we would expect any compressed or shocked region to be orthogonal to the observed orientation, as suggested by numerical simulations of merging clusters (Roettiger et al. \cite{roettiger97}; Ritchie \& Thomas \cite{RT02}, Ricker \& Sarazin \cite{RS01}; Teyssier \cite{teyssier02}). In fact, we can compare our surface brightness distribution and temperature map (Fig. \ref{fig:tmap}) directly with the simulations by Ricker \& Sarazin (\cite{RS01}) and in particular with the bottom set of panels in their Fig.~ 7. These simulations describe a 1:3  mass ratio off-axis merger with an impact parameter $b$ of 5$r_s$ (where $r_s$ is the scale radius of the NFW profile describing the density distribution of the main cluster). Our observed  cluster surface brightness distribution resembles closely, but not exactly, the predicted X-ray surface brightness distribution 1 Gyr after maximum luminosity. At the same time, our temperature map resembles their emission-weighted temperature map at maximum luminosity (0 Gyr). 

We can estimate the time since closest core passage by further comparison with the simulations of Ricker \& Sarazin, taking the above mentioned 1:3 mass ratio, $b=5r_s$ merger case. In Sect \ref{sec:discrete}, with the adoption of the $\Lx -\Tx $ relation of Arnaud \& Evrard (\cite{AE99}) for nearby clusters, we found that the cluster is roughly 2 times more luminous for its temperature. Let us consider the plot of X-ray luminosity versus time in the Ricker \& Sarazin's Fig. 8, taking the {\em quiescent} luminosity to be the value 4 Gyr after maximum. The simulated cluster is twice as luminous at a time $\sim 0.4$ Gyr after maximum luminosity. Note that the average emission-weighted temperature of the simulated cluster at this time has already dropped to approximately the {\em quiescent} temperature.

An alternative estimate of the merger age can be obtained by assuming that the NW edge of the bar is the shock front associated with the off-axis collision between the main cluster and the sub-cluster associated with BG2.
In the Ricker \& Sarazin simulations, the shock forms when the outer regions of the the objects start to interact (their Fig.~6). Let us make the simple assumption that the shock front initially formed at the bottom (South) edge of the hot bar we observe today. Then we can speculate (assuming interaction in the plane of the sky) that the shock front has travelled a distance equivalent to the length of the hot bar, i.e., about 900 $h_{75}$ $^{-1}$ kpc.  The pre-shock gas detected around BG2, in the northern outskirts of the hot bar, is at a temperature k$T_1$=4.8 keV. The shocked gas at the NW edge of the hot bar is at k$T_2$= 7.8 keV (here we adopt the value obtained via the temperature map), yielding a Mach number for the shock of $M$=1.6. From the Mach number and considering that the shock is moving at the speed of the gas behind the shock itself, we calculate that the shock is moving at $v \sim 1800$ km $s^{-1}$. Combining this result with the length of the bar, we find that the shock front has travelled 0.9 $h_{75}^{-1}$ Mpc at a speed of 1800 km $s^{-1}$, implying a merger timescale of $t \lesssim 0.5$ Gyr. This rough estimate is in very good agreement with the result discussed above.

\begin{figure}
\begin{center}
\includegraphics[scale=0.4,angle=0,keepaspectratio,width=\columnwidth]{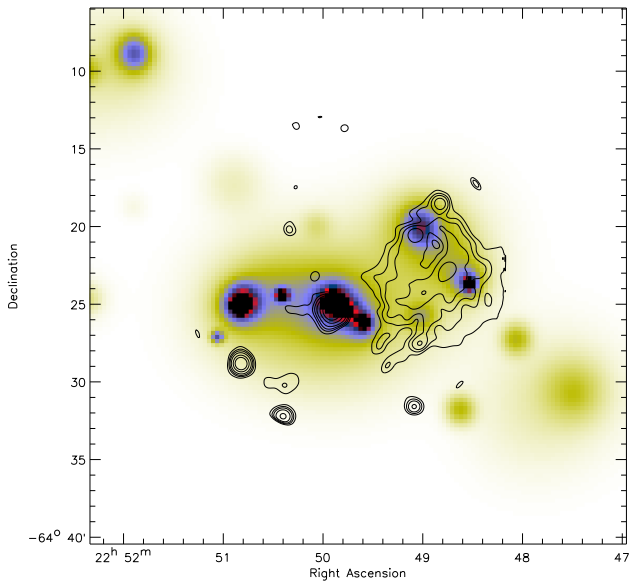}
\caption{X-ray residuals above a 2D-\betamod\ as in Figure \ref{fig:resid2D} overlaid on the red sequence galaxy isodensity map with a cut of 18 in the I-band magnitude (from \cite{f04}. The figure shows that the secondary X-ray peak is located behind the galaxy overdensity in the supposed direction of motion. See text for more details.[\em{see the electronic version of the paper for a colour figure}]}
\end{center}
\label{fig:resopt}
\end{figure}

We thus suggest that we are observing an off-axis merger between two unequal mass objects, where subcluster $A$ (the main cluster) represents the more massive component and subcluster $B$ (around BG2), having come from somewhere in the SSE direction (where we observe the beginning of the hot bar), has already passed beyond closest core passage and is progressing  on an outgoing trajectory towards the NW. This is a natural explanation for the observed surface brightness and temperature structure. Our interpretation is further bolstered by the lack of observable merger signatures in the radial velocity distribution, which would seem to suggest that the motion is taking place in the plane of the sky (see F04 for details). If subcluster B has already passed through the atmosphere of subcluster A, we would expect that the effect of the interaction on the gas would be stronger for the less massive component. In this scenario, it is not surprising that we observe a clear offset between the centroid of the X-ray residuals (Fig.~\ref{fig:resopt}) and the galaxy density peak around BG2, the latter being in advance with respect to the gas in the direction of motion. This would be expected if the galaxies follow the dark matter and the gas behaves as a collisional fluid (see also \cite{f04}). 

Further evidence for our interpretation comes from the ``hot spot'' centred at (J2000) RA = $22^h49^m35.8^s$, Dec = $-64^\circ30\arcmin25\farcs35$  (discussed in Sect.~\ref{sec:tmap}). We can speculate that this is the reverse shock, which can weakly be seen in the lower middle panel of Ricker \& Sarazin's Fig.~7. 

\section{Summary and conclusions}

The main results of our detailed analysis of the \xmm\ observation of the merging cluster A3921, at $z=0.094$ can be summarised as follows:

\begin{itemize}

\item The X-ray morphology of the cluster is elliptical, with a pronounced extension toward the NW, in the direction of the second and third-brightest galaxies (BG2 and BG3). The centroid of the X-ray emission is offset from the brightest cluster galaxy (BG1) by $17\arcsec$ to the NW.

\item A 2D \betamod\ fit excluding all of the extension to the NW shows that the ellipticity of the main cluster is $e = 0.24$. Subtraction of this model from the cluster image reveals a large residual structure in the direction of the extension. This residual contains both diffuse emission and emission from BG2 and BG3. There is further significant residual emission at the centre of the main cluster.

\item The global temperature estimate (k$T = 4.4 \pm 0.07$ keV) is in agreement with previous GINGA and ROSAT results. However, the fit is poor (reduced $\chi^2 = 1.34$) and there are large residuals around 1 keV, suggesting the presence of significant amounts of multi-temperature gas. 

\item The temperature map shows a distinctly nonuniform temperature distribution.  The temperature structure of the Western side is dominated by a ``bar'' at a significantly higher temperature. The hot bar is aligned roughly SE-NW. The Eastern side of the cluster  shows much less temperature variation.

\item A mass profile, calculated under the usual assumptions of hydrostatic equilibrium and spherical symmetry from the surface brightness and temperature profiles of the relaxed-looking Eastern side, is not well fitted by any of the standard profiles from numerical simulations. The best-fitting NFW profile yields $\chi^2_{\nu} > 4$. On the other hand, the optical and X-ray mass estimates agree statistically at a radius of $0.35 h_{75}^{-1}$ Mpc. 

\item Emission from the brightest BG2 and BG3 is clearly visible in both the raw image and in the residual after subtraction of the 2D \betamod. The X-ray temperature of BG3 is rather high k$T = 1.1^{+0.2}_{-0.2}$, and we speculate that it may be the centre of a small group of galaxies. The X-ray emission from BG2 may come from a weak AGN.

\end{itemize}

Comparison of the X-ray results with numerical simulations  leads us to conclude that in A3921 a small subcluster associated with BG2 is undergoing an off-axis merger with a larger main cluster. The merger age appears to be $t \lesssim 0.5$ Gyr. This is in excellent agreement with the optically-derived results of  \cite{f04}.

This analysis of the \xmm\ observation of A3921 has allowed us to completely revise the conventional view of this interesting cluster, which appears currently to be the best example of an X-ray observed large-scale temperature asymmetry.

The present observations illustrate the great advantage to be gained in the study of merging clusters from combining observations in several different wavelengths as well as with numerical simulations. Further spectroscopic observations will give deeper insights into the distribution of galaxies in the subcluster. At the same time, the upcoming {\it Chandra\/} observation of this cluster should allow detailed investigation of the central regions.

\begin{acknowledgements}
We are grateful to M. Arnaud, J-P. Chi\`eze and R. Teyssier for their
participation in the scientific discussion throughout this work. 
We are grateful to C. Ferrari, C. Benoist, and S. Maurogordato for 
sharing with us their optical results,
and we acknowledge the Programme National de Cosmologie (PNC) for
supporting the collaboration between the Observatoire de la C\^ote d'Azur
and the Service d'Astrophysique, CEA-Saclay. We thank
M. Hardcastle for providing the radio map, and M. Birkinshaw and J.P.
Henry for their interesting suggestions.  We thank the 
referee for useful comments and suggestions which improved the manuscript.
E.B. acknowledges the support of PPARC. G.W.P. acknowledges the support
of the European Commission through a Marie Curie Intra-European Fellowship
under the FP6 programme (Contract No.MEIF-CT-2003-500915).

The paper is based on observations obtained with XMM-Newton, an ESA
science mission with instruments and contributions directly funded by ESA
Member States and the USA (NASA). This research has made use of the SIMBAD
database, operated at CDS, Strasbourg, France and of NASA's Astrophysics
Data System.
\end{acknowledgements}


\appendix\label{appx:appx1}

\section{Analysis of systematic effects in temperature determination}

In this Appendix we are interested in assessing possible systematic effects on the global properties of the cluster.

\subsection{$N_H$ variations}

In order to have better constraints on temperature when building the temperature map, we first extracted a spectrum in the region of the cluster less perturbed by the interaction (i.e., by excluding the residuals to the West; this is the {\em main} region as shown in Fig. {\ref{fig:spregions}). We fitted this spectrum with  both an absorbed {\sc mekal} model and an absorbed {\sc apec} model, where  temperature, metallicity\footnote{We adopted the abundances from Grevesse \& Sauval (1998).} and normalisations  (emission measures) were the free parameters of the fit. The Galactic column density was also left as a free parameter. However, our previous experience with \xmm\ data suggests that it is prudent to check for a possible soft excess, which can lead to different temperature values in MOS and pn because of their different sensitivities at low energy.  We performed several tests and concluded that in order for the fitted MOS and pn temperatures to agree within the 90 per cent errors the channels below 0.3 keV for MOS and below 0.5 keV for the pn camera should be excluded. 

The best-fitting results obtained with {\sc mekal} and {\sc apec} are statistically equivalent. The $N_{\rm H}$ value used for the temperature map computation is obtained by simultaneously fitting the MOS spectra only (between 0.3 and 10.0 keV)  with the absorbed {\sc mekal} model. This gives $N_{\rm H}$ = 2.19$_{-0.41}^{+0.42} \times 10^{20}$ cm$^{-2}$ (90 per cent errors for one significant parameter), which is $\sim$ 25 per cent lower than the Galactic value given by Dickey \& Lockman (1990), but in excellent agreement with the PSPC result of $N_{\rm H} = (2.2\pm0.2) \times 10^{20}$ cm$^{-2}$ (Arnaud et al.~\cite{arnauda3921}).

We then investigated possible variations of the N$_{\rm H}$ in the field of view by inspecting the 100$\mu$m map at the position of A3921. Adopting the relation between the infrared emission at 100$\mu$m and the column density of Boulanger et al. (1996), we found values of between 1.93$\times10^{20}$ cm$^{-2}$ and 2.48$\times10^{20}$  cm$^{-2}$, i.e., a maximum variation of 12 per cent. These values are consistent within the errors with our best-fitting X-ray value.  In view of the good agreement between all of these measures, throughout the presented work, the absorption was fixed to the best-fitting value derived from the MOS spectral fit.


\subsection{Global analysis}
We extracted a spectrum in a large circle of radius 11\arcmin, corresponding to $\sim 0.65~r_{200}$, the {\em global} region in Fig. \ref{fig:spregions}. The spectra of the three cameras were then fitted simultaneously with several absorbed thermal models (see Table \ref{tab:spglobmodels}) in the [0.3/0.5 -10.0] keV range (MOS/pn). The free parameters of the fits were the temperature and normalisation  (emission measure). We also investigated whether constraints could be put on abundances of individual elements, which also help in evaluating possible systematics due to specific element variations. We first fitted with all elements as free parameters. However, fitting a {\sc vmekal} or a {\sc vapec} model to  the high S/N global spectrum, we were only able to constrain the relative abundances of O, Ne, Si, Fe, and Ca. Thus in the fits described below we tie all unconstrained elements to  Fe. The Galactic column density was fixed to 2.19$\times10^{20}$ cm$^{-2}$, as discussed above.

\begin{table*}
\begin{center}
\caption{Best-fitting results of the global spectral analysis.}
\label{tab:spglobmodels}
\begin{minipage}{12cm}
\begin{tabular}{l|rrrr}
\hline
single  temperature & & & & \\
\hline
\hline
Parameter \footnote{Errors are quoted at $1 \sigma$ for one interesting parameter}& mekal & apec & vmekal & vapec \\
\hline
kT (keV)& 4.38$ (\pm0.04)$&4.36$^{+0.05}_{-0.07}$&4.40$^{+0.05}_{-0.05}$&4.39($\pm0.06$)\\
Fe\footnote{\footnotesize Abundances are the standard photospheric abundances from Grevesse \& Sauval (1998). For the {\sc mekal} and {\sc apec} models Fe means total abundances. }	& 0.26$ (\pm0.02)$&0.30$ (\pm0.02)$&0.26$ (\pm0.02)$&0.31$ (\pm0.03)$\\
O	& ---			&---			&0.50$ (\pm0.10)$&0.77$^{+0.15}_{-0.14}$\\
Ne	& ---			&---			&0.53$^{+0.11}_{-0.11}$&0.82$^{+0.14}_{-0.14}$\\
Si	& ---			&---			&0.14$ (\pm0.06)$&0.14$ (\pm0.07)$\\
Ca	& ---			&---			&1.14$^{+0.31}_{-0.31}$&1.03$^{+0.32}_{-0.32}$\\
Norm\footnote{The normalisation is in units of [$10^{-14}/ (4\pi (D_A\times(1+z))^2$)] cm$^{-5}$, where D$_A$ is the angular size distance to the source (cm), as defined in the {\sc mekal} model, in XSPEC.
} ($\times10^{-2}$)& 1.86$^{+0.02}_{-0.01}$&1.86$ (\pm0.01)$& 1.83$^{+0.03}_{-0.02}$&1.79$ (\pm0.02)$ \\ 
$\chi^2$/d.o.f. &1551.1/1140&1549.9/1140& 1523.2/1136 & 1511.9/1136 \\
\hline
\hline
two temperatures & & & & \\
\hline
\hline
Parameter & mekal+mekal & apec+apec & vmekal+mekal & vapec+apec \\
\hline
kT1 (keV)& 5.18$^{+0.11}_{-0.10}$&5.09$^{+0.10}_{-0.09}$&5.18$^{+0.11}_{-0.11}$&5.08$^{+0.09}_{-0.11}$\\
Fe$^b$& 0.39$^{+0.03}_{-0.03}$&0.44$^{+0.03}_{-0.03}$	&0.39$^{+0.03}_{-0.03}$&0.44$^{+0.03}_{-0.03}$\\
O	& ---			&---			&0.21$^{+0.12}_{-0.15}$&0.28$^{+0.23}_{-0.21}$\\
Ne	& ---			&---			&0.25$^{+0.14}_{-0.14}$&0.52$^{+0.18}_{-0.18}$\\
Si	& ---			&---			&0.40$^{+0.08}_{-0.08}$&0.45$^{+0.09}_{-0.09}$\\
Ca	& ---			&---			&0.80$^{+0.38}_{-0.38}$&0.87$^{+0.40}_{-0.39}$\\
Norm1$^{c}$($\times10^{-2}$) & 1.70$^{+0.02}_{-0.02}$&1.69$^{+0.02}_{-0.02}$& 1.71$^{+0.02}_{-0.02}$&1.70$^{+0.03}_{-0.02}$ \\ 
kT2 (keV)& 0.65$^{+0.02}_{-0.02}$&0.68$^{+0.03}_{-0.03}$&0.65$^{+0.03}_{-0.03}$&0.67$^{+0.04}_{-0.03}$\\
Fe$^b$& 0.17$^{+0.07}_{-0.04}$&0.17$^{+0.08}_{-0.04}$	&0.16$^{+0.06}_{-0.04}$&0.14$^{+0.09}_{-0.03}$\\
Norm2$^{c}$ ($\times10^{-2}$)& 0.14$^{+0.04}_{-0.04}$&0.15$^{+0.04}_{-0.04}$& 0.16$^{+0.04}_{-0.04}$&0.16$^{+0.04}_{-0.05}$ \\ 
$\chi^2$/d.o.f. &1256.1/1136 & 1263.7/1136 & 1252.6/1132 & 1261.9/1132 \\
\hline
\end{tabular}
\\

\end{minipage}
\end{center}
\end{table*}

The best-fitting results are listed in Table \ref{tab:spglobmodels}. Errors in the Table are quoted at 1$\sigma$ for one significant parameter. All model fits give best-fitting temperature, abundance and normalisation values which agree within their 90 per cent errors. In most cases (the temperature for example), the best-fitting values agree within the 1 $\sigma$ errors. In particular, we note that the derived temperature is not sensitive to abundance variations. The best-fitting single-temperature model is {\sc vapec}, although the reduced  $\chi^2$= 1.3 is clearly not satisfactory.

With the single temperature fit to the global spectrum, we observe large residuals corresponding to the Fe L multiplet (see the top panel of Fig. \ref{fig:spglobfit}), reflecting the existence of multi-temperature gas as observed in the temperature map (Fig. \ref{fig:tmap}).  The largest residuals are centred at $\sim 0.8 keV$, and thus we might expect to observe at least another component at $\sim$ 0.6 keV (Belsole et al. 2001, B\"ohringer et al. 2002). We thus fitted the spectrum with two-temperature thermal models, obtaining  significantly better fits in all models, as shown in Table \ref{tab:spglobmodels}). 

In our two-temperature analysis, we observe the following:
\begin{itemize}
\item we indeed find the 0.6 keV component as expected, although its normalisation is about a factor of 10 lower than the higher temperature component;

\item the best fit is obtained with two {\sc mekal} rather than two {\sc apec} models, which is the opposite of what we  obtained for the one-temperature analysis above; 

\item in both one and two-temperature model fits, the {\sc apec} model gives higher Fe abundances and lower temperatures than the {\sc mekal} model (although we note once more that all results are statistically in agreement within 1$\sigma$). This is likely to be a systematic effect (although a different effect from that found by Buote et al. \cite{buotetal03}, for example). However, because of the resulting $\chi^2$ for the 1T and 2T models, it is difficult to estimate to what extent the temperature/metallicity variation is due to fitting a multi-temperature gas with a single-temperature model, or due to differences in the particular ({\sc mekal, apec})  plasma code;

\item the dominant temperature and the Fe and Si abundances are higher in the two-temperature fits as compared to the single-temperature fits; the opposite is true for O, Ne and Ca. These effects are known as the Fe bias and the Si/S bias (see Buote \cite{buoteabund} for details), and they underline the complexity of estimating metallicity variations in the presence of a multi-temperature gas;

\item in the case of single {\sc vmekal} and {\sc vapec} model fits, the relative abundances of O, Ne, Si and Ca are different at 90 per cent confidence from the relative Fe abundance. Adopting the best-fitting two-temperature model, however, the relative abundance of all elements are consistent within 1 $\sigma$ with the relative abundance of Fe. This result is also partially related to the Fe and Si/S biases discussed in Buote (2000). In view of this result, and given the asymetric temperature distribution of the {\em global} region, it is difficult to conclude if the relative abundances of the elements above are truly different.
\end{itemize}

\noindent These tests suggest that: 
\begin{enumerate}
\item variations in individual abundances have a negligible effect on the temperature determination;
\item neither temperature nor abundance results are strongly dependent on the chosen model.
\end{enumerate}

The existence of the low-temperature model is not related to the central galaxy, as discussed in Sect. \ref{sec:spglob}. This supports our hypothesis concerning the influence of the low-temperature component on the relative abundance of elements other than Fe. In light of the above we conclude that single element metallicity variation is beyond the scope of this paper and anyway is highly uncertain.
\end{document}